\documentclass[aps, prd, twocolumn, superscriptaddress, nofootinbib]{revtex4}
\usepackage{graphicx}
\usepackage{dcolumn}
\usepackage{amssymb}
\usepackage{amsmath,amssymb,amsfonts}

\usepackage{hyperref}
\hypersetup{
   colorlinks   =  true,
    linkcolor    = cyan,
    citecolor    = red,
     urlcolor	=magenta,     
}

\begin{document}

\newcommand{\Eq}[1]{\mbox{Eq. (\ref{eqn:#1})}}
\newcommand{\Fig}[1]{\mbox{Fig. \ref{fig:#1}}}
\newcommand{\Sec}[1]{\mbox{Sec. \ref{sec:#1}}}

\newcommand{\PHI}{\phi}
\newcommand{\PhiN}{\Phi^{\mathrm{N}}}
\newcommand{\vect}[1]{\mathbf{#1}}
\newcommand{\Del}{\nabla}
\newcommand{\unit}[1]{\;\mathrm{#1}}
\newcommand{\x}{\vect{x}}
\newcommand{\y}{\vect{y}}
\newcommand{\p}{\vect{p}}
\newcommand{\ScS}{\scriptstyle}
\newcommand{\ScScS}{\scriptscriptstyle}
\newcommand{\xplus}[1]{\vect{x}\!\ScScS{+}\!\ScS\vect{#1}}
\newcommand{\xminus}[1]{\vect{x}\!\ScScS{-}\!\ScS\vect{#1}}
\newcommand{\diff}{\mathrm{d}}
\newcommand{\mk}{{\mathbf k}}
\newcommand{\ep}{\epsilon}

\newcommand{\be}{\begin{equation}}
\newcommand{\ee}{\end{equation}}
\newcommand{\bea}{\begin{eqnarray}}
\newcommand{\eea}{\end{eqnarray}}
\newcommand{\vu}{{\mathbf u}}
\newcommand{\ve}{{\mathbf e}}
\newcommand{\vn}{{\mathbf n}}
\newcommand{\vk}{{\mathbf k}}
\newcommand{\vp}{{\mathbf p}}
\newcommand{\vx}{{\mathbf x}}
\newcommand{\PP}{{\mathbb P}}
\def\dup{\;\raise1.0pt\hbox{$'$}\hskip-6pt\partial\;}
\def\ddn{\;\overline{\raise1.0pt\hbox{$'$}\hskip-6pt\partial}\;}


\title{Life of cosmological perturbations in MDR models, and the prospect of travelling primordial gravitational waves}

\newcommand{\addressBurgos}{Departamento de F\'isica, Universidad de Burgos, E-09001 Burgos, Spain}
\newcommand{\addressIC}{Theoretical Physics, Blackett Laboratory, Imperial College, London, SW7 2BZ, United Kingdom}

\author{Giulia Gubitosi}
\affiliation{\addressBurgos}
\author{Joao Magueijo}
\affiliation{\addressIC}

\date{\today}

\begin{abstract}
We follow the life of a generic primordial perturbation mode (scalar or tensor) subject to modified dispersion relations (MDR), as its proper wavelength is stretched by expansion. A necessary condition ensuring that  travelling waves can be converted into standing waves is that the mode starts its life deep inside the horizon and in the trans-Planckian regime, then leaves the horizon as the speed of light corresponding to its growing wavelength drops, to eventually become cis-Planckian whilst still outside the horizon, and finally re-enter the horizon at late times. We find that  scalar modes in the observable range satisfy this condition, thus ensuring the viability of MDR models in this respect. For tensor modes we find a regime in which this does not occur, but in practice it can only be realised  for wavelengths in the range probed by future gravity wave experiments if the quantum gravity scale experienced by gravity waves goes down to the PeV range. In this case travelling---rather than standing---primordial gravity waves could be the tell-tale signature of MDR scenarios.

\end{abstract}


\maketitle

\section{Introduction}

Models of the primordial universe where perturbations satisfy Planck-scale-modified dispersion relations (MDR) have been shown to reproduce a number of observed properties of the power spectrum without resorting to an inflationary phase of expansion. The simplest of such models, where the dispersion relation reads:
\be
E^{2}=p^{2}\left(1+(\lambda p)^{2\gamma}\right) \label{eq:MDR}
\ee
can produce a (quasi) scale-invariant power spectrum, with amplitude depending on the ratio between the deformation parameter $\lambda$ and the Planck scale, and spectral index depending on the value of the dimensionless parameter $\gamma$~\cite{Amelino-Camelia:2013tla,Amelino-Camelia:2013wha,Amelino-Camelia:2013gna,Amelino-Camelia:2015dqa}. Moreover, once scalar perturbations   exit the horizon  they  subsequently re-enter it  as standing waves with the correct temporal phase \cite{Gubitosi:2017zoj, Gubitosi:2017jwi}, so that the position of the Doppler peaks in the spectrum matches observations. This is an important non-trivial detail, amounting to examining the equivalent of ``squeezing'' in inflationary models for these alternative scenarios.

On a different front, it is interesting that these models are motivated by quantum gravity theories predicting
that the dimensionality of space-time runs with the energy, approaching dimension 2 in the UV. This is  
true in Ho\v{r}ava-Lifshitz (H-L) gravity \cite{Horava:2009if}, but similar behaviour was discovered in 
other approaches, such as the renormalization-group analysis leading to the  ``asymptotic-safety approach"~\cite{Krasnov:2012pd,Lauscher:2005qz}. Hints in the same direction (perhaps more elusively
related to MDRs and the work on fluctuations) also arise in  
Loop Quantum Gravity \cite{Modesto:2008jz} and Causal Dynamical Triangulations \cite{Ambjorn:2005db}.

The basic assumption justifying the relevance of these quantum gravity models for primordial cosmology is that the  perturbation scales that are within the observable range nowadays correspond to (super-) Planckian scales at the time the perturbations were produced. Thus, the dispersion relation (\ref{eq:MDR}) is dominated by the UV limit, $E\simeq p\left(\lambda p\right)^{\gamma}$, producing the interesting observable consequences reviewed in Section \ref{sec:Constraints} (where we evaluate the range of model parameters that produce a power spectrum with the observed amplitude and spectral index).
This assumption seems reasonable, because, going back in time, the scale factor $a$ eventually shrinks enough so as to make the physical momentum $p=\frac{k}{a}$ associated to any given comoving (and thus fixed)  wavenumber, $k$, enter the super-Planckian regime (in the inflationary scenario this fact  causes the so-called trans-Planckian problem  \cite{Martin:2000xs, Brandenberger:2012aj}). 

Currently observable scales correspond to extremely small wavenumbers. The largest wavelength that we can observe (corresponding to the smallest observable wavenumber $ k_\text{min}$)  is equal to the Hubble radius today:\footnote{In this paper we set the value of the scale factor today $a_{0}=1$ and we use  units such that $c=\hbar=G=1$.}
\be
k_\text{min} = a_{0}H_{0}\simeq 10^{-32} \,\text{eV}.
\ee
The Cosmic Microwave Background (CMB) power spectrum probes  wavelengths that are up to $\sim 4$ orders of magnitude smaller (i.e. $10$ e-foldings). Therefore, any theory of the early universe must be applicable to wavenumbers in the range:
\bea
a_{0} H_{0}< &k& < 10^{4} a_{0} H_{0} \label{krange1}\\
\nonumber\\
\Rightarrow 10^{-32}  \,\text{eV}< &k& < 10^{-28}  \,\text{eV}\,. \label{krange2}
\eea
It is then clear that these modes are in the MDR phase (i.e. their dispersion relation is dominated by the UV correction) during {\it very} early stages of the universe, since the scale factor must contract by at least $60$ orders of magnitude in order to make their physical momentum Planckian. At these times background quantities, such as the Hubble rate and temperature, may, of course, take extreme values, which we  compute in this paper.

The above considerations are based on the assumption that while the perturbations are subject to Planck-scale corrections, the background behaves as in standard general relativity. This assumption is generally made in MDR models, and  is also fundamental for the work presented here. In particular, in this framework one still has a big-bang singularity, unless further quantum gravity effects enter into the picture at even earlier times than those relevant for this work.

\begin{figure}[h]
\includegraphics[scale=0.4]{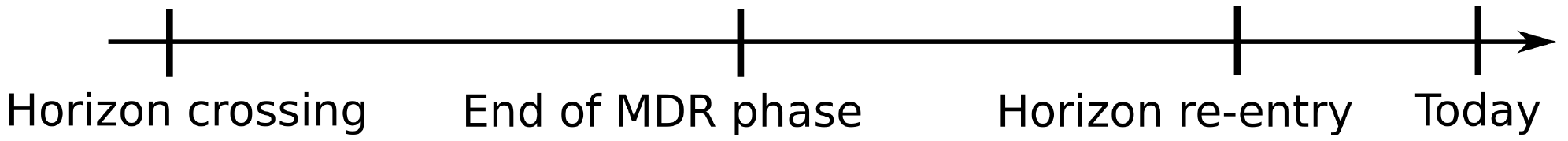}
\caption{Qualitative depiction of the timeline of the relevant cosmological epochs for a typical fixed comoving mode. \
The scale factor (and so time, assuming an expanding Universe) increases from left to right.  }
\label{fig:timeline0}
\end{figure}

As we mentioned, the reliability of the MDR model is grounded on the fact that scalar modes exit the horizon in the MDR phase, and then spend enough time outside the horizon so as to turn into standing waves before they re-enter the horizon during the standard (i.e. non-MDR) epoch (see the qualitative time-line in Figure \ref{fig:timeline0}). In \cite{Gubitosi:2017jwi} we focussed on the amount of time the modes  spend outside of the horizon in the MDR model, assuming they do exit the horizon in the MDR phase.   This assumption is put under scrutiny in this paper, where we investigate the conditions that allow the modes to actually exit the horizon while in the MDR regime. We will show that MDR cosmology can indeed comply with the required picture for scalar modes in the CMB observable range. For tensor modes we still lack observational probes of the properties of perturbations as they re-enter the horizon. They could be standing or travelling waves, for all we know (assuming they exist at all). 
In this paper  we will show that tensor modes falling in the future gravity waves detectors range might form travelling waves, if the deformation scale for tensor modes, $\lambda_T^{-1}$, is as low as the PeV scale. We will argue that this could provide  an observational signature of MDR models, distinguishing them from alternatives, but only if further theoretical arguments 
can be found supporting such a low scale for the onset of quantum gravity effects (as has been suggested, starting from~\cite{ArkaniHamed:1998rs}).
 
The plan of this paper is as follows. Having reviewed the subject in Section \ref{sec:Constraints}, 
in Section \ref{sec:HorizonProblem} we show that the requirement that the horizon problem be solved within MDR models implies a consistency relation between the power law parameter $\gamma$ and the equation of state during the MDR phase. 
The values of the scale factor associated with the relevant cosmological epochs (marked in the timeline of Figure \ref{fig:timeline0}) are computed in Section \ref{sec:CosmoPhases}. With increasing scale factor the modes  transition from the MDR phase, where the UV correction to the dispersion relation dominates, to a standard cosmology phase, where the dispersion relation is the usual $E^{2}=p^{2}$. 
If the end of the MDR phase happens before the modes actually exit the horizon, then they can never do so---the (quasi) scale invariance of the power spectrum is spoiled, and moreover perturbations do not form standing waves. 

The conditions to avoid this are also discussed in Section \ref{sec:CosmoPhases}.
We find that scalar modes within the CMB  observable range follow indeed the required timeline, and are thus allowed to form standing waves. Moreover we compute the range of wavelengths for which travelling waves could be observed in the tensor perturbations sector, were primordial gravity waves to be detected by  forthcoming experiments.

Finally, implications  are reviewed in a concluding Section.

\section{Observational constraints on MDR models}\label{sec:Constraints}
Depending on which frame (or representation) is chosen, the gravity affecting the perturbations may be
rainbow gravity~\cite{Magueijo:2002xx,Amelino-Camelia:2013wha} or the standard one~\cite{Amelino-Camelia:2013tla} with underlying implications examined in~\cite{Amelino-Camelia:2015dqa}. 
Here we choose
the so-called MDR frame, so that the second-order action for the primordial perturbations is the same as in the standard  general relativity (GR) theory. Then, the equation of motion for the Fourier mode $v_{k}$ is:
\be
v_{k}''+\left[c(k,\eta)^{2} k^{2}-\frac{a''}{a}\right]v_{k} = 0\,, \label{EOM}
\ee
where $'$ denotes derivative with respect to conformal time $\eta$, related to comoving time $t$ by $d\eta = a(t) dt$. The comoving gauge curvature perturbation is given by $v$ via $\zeta=-\frac{v}{a}$.
Because perturbations obey  the modified dispersion relation \eqref{eq:MDR}, the velocity  $c=\frac{ E}{ p}$ depends on the wavenumber and the scale factor:
\be
c(k,a)= 
\sqrt{\left(1+\left(\lambda {k}/{a}\right)^{2\gamma}\right)}\,.
\ee

Consistently with the assumption that the background behaves as in classical GR, we assume a power-law dependence of the scale factor on conformal time:
\be
a(\eta)/a(\eta_*)=\left(\eta/\eta_*\right)^{m}\,,\label{scalefactor}
\ee
where $\eta_*$ is some reference time during the MDR phase\footnote{After the MDR phase we  assume that the universe behaves as in the standard $\Lambda$CDM model, see section  \ref{MDRStandardTransition}.} and as usual $m=\frac{2}{1+3w}$ in terms of the (constant) equation of state parameter $w$ dominating the background evolution durng the MDR phase. We assume $m>0$, consistent with the choice motivated in the following Section \ref{sec:HorizonProblem}.

It has been established  \cite{Amelino-Camelia:2013tla}  that for $\gamma=2$ the spectrum of perturbations is scale-invariant, regardless of the equation of state. In order to see this one can compare the solutions to the equation of motion \eqref{EOM} inside and outside the horizon. While modes are inside the horizon the first term in brackets in \eqref{EOM} dominates.\footnote{The following Section demonstrates that in MDR models the horizon problem is indeed solved, so that modes start off inside the horizon and subsequently exit.} Then the normalised vacuum solution is given, up to a phase, by \cite{Amelino-Camelia:2013tla}:
\be
v_{k}\sim \frac{1}{\sqrt{c(k,a) k}} = \sqrt{\frac{a^{\gamma}}{\lambda^{\gamma}k^{\gamma+1}}}\,,
\ee
where we used the UV value of the velocity,
\be
c=\left( \frac{\lambda k}{a}\right)^{\gamma}\,.\label{UVc}
\ee
This solution has to be matched to the one outside the horizon, when the second term in  brackets in \eqref{EOM} dominates. For this solution one can make the ansatz:
\be
v_{k}\sim F(k) a\,,
\ee
where the function $F(k)$ is determined by the matching condition at horizon crossing.
It is immediate to see that for $\gamma=2$ the power spectrum $P_{\zeta}(k)\sim k^{3}|v_{k}|^{2}$ is already scale invariant inside the horizon. Given that in this case the dependence on the scale factor is the same inside and outside the horizon, the scale invariance is maintained at horizon crossing, independently of the equation of state.

However, because the  power spectrum is observed to be slightly red, $P_{\zeta}(k)\sim k^{n_{S}-1}$ with spectral index  $n_{S}\simeq 0.96$ \cite{Planck:2013jfk}, one must require \cite{Amelino-Camelia:2013tla} that $\gamma\lesssim 2$ at the time perturbations cross the horizon.\footnote{This can either be the actual value of $\gamma$, or some effective value at an intermediate sub-UV regime \cite{Amelino-Camelia:2013tla}.}  In this case however the scale factor does not cancel out anymore when matching the modes at horizon crossing, so that a dependence on the equation of state parameter $w$ is introduced. In fact, the matching condition for general values of $\gamma$ reads:
\be
 \sqrt{\frac{a^{\gamma}}{\lambda^{\gamma}k^{\gamma+1}}}= F(k) a,\label{matching}
\ee
to be computed at horizon crossing, that is when the two terms in brackets in \eqref{EOM} have equal magnitude:
\be
c(k,a)^{2} k^{2}=\left|\frac{a''}{a}\right| \quad\Rightarrow \quad  k^{\gamma+1} \propto a^{\gamma -1/m}\,.
\ee
Plugging this into \eqref{matching} one obtains:
\be
n_{S}-1 = \frac{(\gamma-2)(m+1)}{1-m\gamma} = \frac{3(\gamma-2)(1+w)}{1+3w-2\gamma},
\ee
which for the observed value of the spectral index $n_{S}$ and reasonable values of the equation of state parameter (see following Section), implies that $\gamma$ falls in the range
\be
\gamma\in (1.98,  2)\,.\label{gammaConstraint}
\ee

The deformation parameter $\lambda$ is constrained by the amplitude of the power spectrum,
\be
A_S=P_{\zeta}(k_*)\simeq 2\cdot 10^{-9}\,,
\ee
at $k_*=0.05 \;\text{Mpc}^{-1}$ \cite{Martin:2013tda}. In fact, the power spectrum depends on the ratio between the energy scale $\lambda^{-1}$ and the Planck energy as follows:
\be
P_{\zeta}(k)\equiv\frac{k^{3}}{2\pi^{2}}\left| \frac{v_{k}}{a M_{P}}\right|^{2} = \frac{k^{2}}{4 \pi^{2}}\frac{1}{a^{2} M_{P}^{2} c} =   \frac{k^{2-\gamma}}{4 \pi^{2}}\frac{a^{\gamma-2}}{ M_{P}^{2}\lambda^{\gamma}}\,,
\ee
in which $M_{P} = 2.4\cdot 10^{27}\,\text{eV}$  is the reduced Planck mass.
Using $\gamma\simeq 2$ one finds:
\be
\lambda^{-1}\simeq 3\cdot 10^{-4}M_{P}\simeq 7\cdot 10^{23}\, \text{eV}\,.\label{ellConstraint}
\ee
So the  dispersion relation of perturbations  is deformed at a scale that is a few orders of magnitude below the Planck scale.

Of course the constraints \eqref{gammaConstraint} and \eqref{ellConstraint} only  apply to the   dispersion relation of scalar perturbations. Even assuming that tensor perturbations  obey a  deformed dispersion relation similar to that of scalar perturbations, its parameters can in principle take different values: \be
E^{2}=p^{2}\left(1+  (\lambda_T\, p)^{2\gamma_T}\right)\,, \label{eq:MDRtensor}
\ee
where the index $T$ denotes parameters referring to tensor modes.
Since tensor perturbations have not been observed yet $\gamma_T$ is unconstrained.
On the other hand, the observational upper bound on the tensor-to-scalar ratio $r<0.07$ \cite{Akrami:2018odb} implies  a lower bound on the deformation parameter $\lambda_T$. In fact, defining the tensor amplitude $A_T$ in a similar way as the scalar amplitude one finds
\be
r\equiv \frac{A_T}{A_S}=\left(\frac{k}{a}\right)^{\gamma-\gamma_T} \lambda^\gamma\lambda_T^{-\gamma_T}\,.
\ee 
Assuming that the velocity of tensor modes has a similar power-law as scalar modes, $\gamma_T\simeq\gamma$:
\be
r\simeq\left(\frac{\lambda}{\lambda_T}\right)^\gamma\,,
\ee
so that 
\be
r<0.07\Rightarrow \lambda_T>4 \lambda\,.
\ee
Note that higher-order corrections to the dispersion relation of gravity waves of the sort we are considering here are very poorly constrained, the best current limit being given by observations of black hole binaries \cite{Sotiriou:2017obf}. So in principle $\lambda_T$ could be several orders of magnitude larger than $\lambda$.

\section{MDR solution to the horizon problem}\label{sec:HorizonProblem}

In order for a model of the primordial universe to be phenomenologically viable,  the scalar modes contributing to the measured spectrum of the CMB must form standing waves with a given temporal phase, and this should be such that it implies the correct position of the Doppler peaks in the observed CMB power spectrum \cite{Dodelson:2003ip, Gubitosi:2017zoj,Gubitosi:2017jwi}.
This is achieved if the modes start off inside the horizon, are pushed outside it and spend there a sufficiently long time before horizon re-entry at late times, so that the momentum of fluctuations is suppressed. This happens in inflationary models, as well as in MDR models producing a scale invariant or red power spectrum  \cite{Gubitosi:2017zoj}. 

Here we  focus on the minimal requirement that in an expanding universe modes in the MDR regime evolve from sub-horizon to super-horizon size, revising the argument in \cite{Amelino-Camelia:2013tla}. This leads to a consistency relation  between the equation of state parameter $w$ and the parameter $\gamma$. In the following Section we investigate the conditions that guarantee that the modes do not exit the MDR regime before horizon crossing, which would spoil the results of this Section.

The correct sub-horizon/super-horizon transition happens if the first term in brackets in Eq.\eqref{EOM} dominates at early times and the second one dominates at late times. The dependence of the two terms on conformal time is:
\bea
&&c^{2}\simeq\left( \frac{\lambda k}{a}\right)^{2\gamma}\sim \eta^{- 2 m \gamma}\,,\nonumber\\
&&\left|\frac{a''}{a}\right|\sim \eta^{-2}\,. \label{comparison}
\eea
To verify that the evolution of these quantities is the correct one we first need to understand how conformal time $\eta$ changes as the universe expands. The condition for the universe to expand is that  $\frac{a'}{a}\equiv m/\eta>0$ (see Eq. \eqref{scalefactor}). This can be  achieved in two ways.  For $m<0$ (i.e. $w < -1/3$)  the conformal time must be negative and approaching zero from $-\infty$. It is easy to see that in this case the horizon problem is solved for any value of $w<-1/3$. This is indeed the scenario where  the universe  undergoes a phase of accelerated expansion as in inflation.

In the rest of the paper we will focus on the more interesting  case $m>0$ (i.e. $w> -1/3$), where the conformal time must be positive, $\eta>0$, and increasing from $0$. In this scenario the first term in \eqref{comparison} decreases faster than the second one only if
\be\label{HorizonSolutionCondition}
m\gamma > 1\,,
\ee 
or equivalently if the equation of state parameter is constrained in the interval 
\be
-\frac{1}{3}< w < \frac{2 \gamma  - 1}{3}\,.\label{wConstraint}
\ee
In this case there is no inflationary-type behaviour, but still the horizon problem is solved thanks to the nonstandard evolution of the modes. This can be related to the fact that  in the rainbow frame, where the dispersion relation is made trivial at the expense of modifying gravity, one actually finds that  there is inflation \cite{Amelino-Camelia:2013wha, Barrow:2013gia, Garattini:2012ca}. In fact, one can show that in the rainbow frame the evolution of the scale factor with time is the same as in standard GR, but with an effective equation of state parameter $\tilde w\equiv w-\frac{2}{3}\gamma$ \cite{Barrow:2013gia}.  So in the rainbow frame the condition \eqref{wConstraint} reads $ -\frac{2\gamma+1}{3}<\tilde w<-\frac{1}{3}$, which implies that there is inflation (in particular, for $\gamma=2$ and $w=\frac{1}{3}$ there is de Sitter inflation in the rainbow frame, $\tilde w=-1$).

As a closing note to this Section, we can look at the $\gamma=0$ limit to understand  the implications for  constant $c$ models.\footnote{However keep in mind that the correct classical limit of the MDR model is achieved for $\lambda\rightarrow 0$.} In this limit the two sides of constraint \eqref{wConstraint} contradict each other, so that in the $m>0$ case one can not solve the horizon problem. The only viable option is then  $m<0$ (i.e. $w<-\frac{1}{3}$). This implies that only inflationary models can solve the horizon problem if $c$ is constant and the universe is expanding.

\section{Horizon crossing and MDR-standard phase transition}\label{sec:CosmoPhases}

In the previous Section we derived the consistency relation between $\gamma$ and $w$ that ensures  perturbations start off inside the horizon and then are pushed outside as the universe expands. The underlying assumption of the analysis is that we can use the  UV value of the velocity $c$ all the way until horizon crossing, see Eq. \eqref{comparison}. Namely, we showed that if  perturbations are in the MDR regime then when the consistency condition \eqref{wConstraint} is satisfied modes transition from   sub-horizon to  super-horizon scales. 

The rationale behind the assumption that modes are in the MDR regime  until they exit the horizon is that, given that in the cosmological model we are considering there is no accelerated expansion, horizon crossing  can indeed  only happen  if perturbations are in the MDR regime. If  a mode transitions to the standard non-MDR regime before exiting the horizon, then it will never do so, disrupting the generation of coherent perturbations at late times. In this Section we scrutinise these matters and derive the conditions that guarantee that modes  stay in the MDR regime at least until horizon crossing, so that in MDR cosmological models perturbations go through the following sequence of relevant events:
\begin{itemize}
\item  horizon exit (during the primordial MDR phase),
\item  transition from the MDR epoch  to standard cosmological evolution,
\item horizon re-entry at late times. 
\end{itemize}
The last of these events happens in the standard cosmology phase, so the condition determining the time of horizon re-entry for each mode $k$ is the usual one:
\be
k=a_\text{re-entry} H_\text{re-entry}\,.
\ee
The first two of these events are instead affected by the modified evolution of perturbations determined by the MDR.
As we discuss in the following, the values of the scale factor corresponding to each of these events depend on the wavenumber $k$. This means that each mode exits the horizon or transitions from the MDR to standard behaviour at a different time.\footnote{A $k$-dependence is already  found in inflationary cosmology as far as  horizon crossing is concerned. In MDR cosmology also the end of the MDR phase is $k$-dependent, as opposed to the end of the inflationary phase, which is governed by the background evolution, so it is universal.} Failure to respect this sequence  of events by a mode with given frequency results in the mode not producing  standing waves.
In the following we evaluate the time at which horizon crossing and MDR-standard phase transition happen as a function of the wavenumber $k$. By comparing the two we can infer the range of frequencies for which the two events happen in the order that allows for the formation of standing waves.

\subsection{Horizon crossing}\label{HorCrossing}

Horizon crossing happens when the two terms in brackets in Eq. \eqref{EOM} have equal magnitude:
\be
c(k,a_{k})^{2} k^{2}=\left|\frac{a_{k}''}{a_{k}}\right|\,.\label{horizonCrossing}
\ee
Clearly this identifies a different scale factor $a_{k}$ for each wavenumber $k$, so that each mode exits the horizon at a different time.

We  proceed under the assumption that horizon crossing happens  during the MDR phase, so that  the velocity $c(k,a)$ takes the UV form of Eq. \eqref{UVc}. This approximation is equivalent to
\be
\left(\frac{\lambda k}{a}\right)^{\gamma}\gg  1\Rightarrow  \frac{k}{a}\gg\lambda^{-1} \,,\label{UVcondition}
\ee
meaning  that  the physical momentum $p_{k}\equiv  \frac{k}{a_{k}}$ is super-Planckian. We will verify \emph{a posteriori} if this condition is satisfied until after horizon crossing.

In the UV regime the condition \eqref{horizonCrossing} for horizon crossing can be written as:
\be
\lambda^{2\gamma} k^{2(1+\gamma)}=\left|a_{k}''\right| a_{k}^{2\gamma-1}.\label{HorizonCrossing}
\ee
Then using  Eq. \eqref{scalefactor}  the (conformal) time $\eta$ at which  horizon-exit happens for a given mode $k$ is found:
\be
\frac{\eta_{k}}{\eta_{*}}=\left(\eta_{*}a(\eta_*)^{-\gamma}\, k^{1+\gamma}  \lambda^{\gamma}  \left|(m-1)m\right|^{-\frac{1}{2}}\right)^{\frac{1}{m\gamma-1}}\;  \label{timeHorExit}\,.
\ee
As we mentioned, $\eta_*$ is a reference time during the MDR phase. In the following section we will take it to be the time at which the MDR phase ends, $\eta_*=\eta_\text{end}$.
Note that for $m=1$ ($w=1/3$) this relation is pathological, regardless of the value of $\gamma$, and in particular for the ``standard'' $\gamma=0$ case. This is because for $w=1/3$ the universe is radiation-dominated. Radiation is conformally coupled to gravity, so that $a''/a=0$ in the equation of motion \eqref{EOM}.\footnote{In the radiation dominated universe perturbations are still sensitive to the background expansion of the universe through their conjugate momentum, which produces squeezing and thus generation of standing waves \cite{Gubitosi:2017jwi}.} As explained at the end of Section \ref{sec:HorizonProblem}, one can deal with the $w=1/3$ case by looking at the rainbow frame, where for $w=1/3$ one has standard de Sitter inflation.

The first modes to exit the horizon are those corresponding to the largest wavelength/smallest wavenumbers $k_\text{min}\sim 10^{-32} \,\text{eV}$, which are re-entering the horizon today.

We can also set a lower bound on the value of the Hubble rate $H=\frac{a'}{a^{2}}$ at horizon crossing. Starting from Eq. \eqref{HorizonCrossing} and using Eq. \eqref{scalefactor} to write $\frac{a''}{a}= m(m-1) \eta^{-2}$ and $a^{2}H^{2} = m^{2}\eta^{-2}$ the condition for horizon crossing becomes:
\be
a_{k}H_{k}^{\frac{1}{(1+\gamma)}}=k \; \lambda^{\frac{\gamma}{(1+\gamma)}}  \left|\frac{m-1}{m}\right|^{-\frac{1}{2(1+\gamma)}} \label{kHkRelation}\,.
\ee
By substituting this into the condition stating that physical momenta are super-Planckian, Eq. \eqref{UVcondition}, one finds that also the Hubble rate must be beyond the Planck scale:
\be
 H_{k} \gg \lambda^{-1}   \left|\frac{m}{m-1}\right|^{\frac{1}{2}}\,. \label{UVconstraint}
\ee
This condition on the  Hubble rate is the same for all modes and does not depend on the specific form of the UV dispersion relation (i.e. there is no $k$ nor $\gamma$ dependence).
Once we compute the time of MDR-standard phase transition $\eta_\text{end}$ in the following section, we will be able to explicitly compute $\eta_k$ from  \eqref{timeHorExit} and thus $a_k$ and $H_k$, verifying whether the above condition is satisfied.

\subsection{MDR-standard phase transition}\label{MDRStandardTransition}

Assuming a sudden transition between the UV MDR phase and the standard cosmology phase, this happens when the speed of propagation of the perturbations becomes standard:
\be
\left(\frac{\lambda k}{a_\text{end}}\right)\approx 1 \Rightarrow a_\text{end} \approx \lambda k\,. \label{MDRendCONDITION}
\ee
Note that the end of MDR phase happens at different times for modes with different wavelengths and it does not depend on the value of $\gamma$.
So for  the range of observable scales included in Eq. \eqref{krange1} the MDR-standard phase transition happens for values of the scale factor included in the interval 
\be
10^{-56} < a_\text{end} < 10^{-52}\,, \label{ENDofMDRscalefactor}
\ee
where we used the observational constraint \eqref{ellConstraint} and
the lower value of the scale factor corresponds to the transition for the lower wavenumber/larger wavelength modes. 

Since in the standard cosmological phase the Hubble parameter evolves as:
\be
H(a)=\sqrt{H_{0}^{2}\left(  \Omega_{r,0} a^{-4} +  \Omega_{m,0} a^{-3} +  \Omega_{\Lambda}  \right)}\,,
\ee
we can compute its value at the end of the MDR phase. For the largest wavelengths, corresponding to $k_\text{min}$, we find:
\be
H_{\text{end}}|_{k_\text{min}} = 10^{78}\,\text{eV} \simeq 10^{50} M_{p}\,,
\ee
using $\Omega_{r,0} = 9\cdot 10^{-5}$, $\Omega_{m,0} = 0.32$ and $\Omega_{\Lambda} = 0.68$. So when the largest wavelengths transition to the standard cosmological phase the Hubble parameter is still much larger than the Planck scale,\footnote{These values would change if a nonstandard phase existed between the end of MDR regime and beginning of radiation-domination.} and the same holds for the temperature $\frac{T_{\text{end}}}{T_{0}} = \frac{a_{0}}{a_{\text{end}}}$:
\bea
 T_{\text{end}}|_{k_\text{min}} =  10^{49}\,\text{eV} \simeq 10^{21}T_{p}\,.
\eea
Smaller wavelengths exit the MDR phase when the Hubble parameter takes smaller values, eventually reaching its current value for Planckian wavelengths, which are exiting the MDR phase today.

From the estimates just discussed we  notice that the MDR phase is characterized by a super-Planckian regime for some background quantities. This might raise concerns about the assumption of a   background evolution governed by standard Einstein gravity. However, we find that the scalar curvature is far from this potentially problematic regime, and has actually a value close to zero. To show this we compute the Ricci scalar $R=6 \dot H(t)+12 H(t)^{2}$, where $t$ is the comoving time. At the end of the MDR phase one finds
\be
\dot H_{\text{end}}=- \frac{ H_{0}^{2} (4  \Omega_{r,0}+3  \Omega_{m,0}a(t))}{2 a(t)^{4}} \simeq  -  2 H_{0}^{2}   \Omega_{r,0}a_{\text{end}}^{-4}\,,
\ee
where in the last step we used the fact that for very small values of $a(t)$ the term proportional to $\Omega_{m,0}$ is negligible.
So the $\dot H$ term in the Ricci scalar cancels with the $H^{2}$ term, which for very small values of $a(t)$  can be approximated as
\be
H_{\text{end}}^{2}\simeq  H_{0}^{2}  \Omega_{r,0} a_{\text{end}}^{-4} \,.
\ee
The fact that the scalar curvature is very close to zero at the end of the MDR phase gives support to the reasonability of the assumption of a standard background evolution, since  what is usually assumed to be the  full quantum gravity regime is that of Planckian curvature.\footnote{On the other hand, new concerns arise if one studies the curvature via the Kretschmann scalar, which is used e.g. in the study of black holes curvature, since the Ricci scalar is zero for the Schwarzschild metric. In a spatially-flat FRW background the Kretschmann scalar reads $K= 12 (\dot H^{2}+2 H^{4}+2 H^{2}\dot H)$, and one can easily verify that this gives $K\simeq 24 H^2$ when evaluated at the end of the MDR phase, using the approximations that apply to the Ricci scalar discussed in the main text. So while the Ricci scalar is zero, the Kretschmann scalar is super-Planckian. The issue of whether this would require quantum effects to be introduced also in the evolution of the background is  left  for future investigation.
}

In terms of the conformal time the end of the MDR phase for a mode $k$ happens when:
\be
\eta_{\text{end}}=\frac{m}{a_\text{end} H_\text{end}}= \frac{m \lambda k}{H_0\sqrt{\Omega_{r,0}}},\label{timeMDRend}
\ee
having used $a H=m\eta^{-1}$, together with Eq. \eqref{MDRendCONDITION} and $H(a)\simeq H_0 a^{-2}\sqrt{\Omega_{r,0}}$ for small values of $a$. 
The first modes to transition to the standard phase are those with the largest wavelength, and they do so at time
\be
\eta_{\text{end}}|_{k_\text{min}} \approx  m \cdot 10^{-22} \,\text{eV}^{-1}  \approx m \cdot10^{-36} \,\text{s}\,.\label{etaendkmin}
\ee

Having computed the time of MDR-standard phase transition as a function of the model parameters and of the  wavenumber $k$, Eq. \eqref{timeMDRend}, this can be used in Eq. \eqref{timeHorExit} to get a more informative expression for the time of horizon exit:
\be
\frac{\eta_{k}}{\eta_{\text{end}}}=\left( \frac{ \lambda k^2}{H_0\sqrt{\Omega_{r,0}}}  m \left|(m-1)m\right|^{-\frac{1}{2}}\right)^{\frac{1}{m\gamma-1}}\;  .\label{eq:etaketaend}
\ee 
This expression holds for both scalar modes and  tensor modes, if one uses the relevant parameters $\lambda_T$ and $\gamma_T$ as explained at the end of Section \ref{sec:Constraints}.

We are now in a position to compare the time of horizon exit with that of MDR-standard phase transition and verify whether the modes of any given wavelength exit the horizon in time.\footnote{Note that the assumption \eqref{UVcondition} is verified as long as $\eta_k/\eta_{\text{end}}<1$.} For both scalar and tensor perturbations Eq. \eqref{eq:etaketaend} indicates that for large enough modes $\frac{\eta_{k}}{\eta_{\text{end}}}<1$, so these modes  will be outside the horizon at the end of the MDR phase, their horizon exit having happened at some earlier time, when they were still in the MDR regime. These are the modes that can form standing waves, if they spend enough time outside the horizon \cite{Gubitosi:2017jwi}. Smaller wavelength/larger wavenumber modes, such that $\frac{\eta_{k}}{\eta_{\text{end}}}>1$, fail to exit the horizon before transitioning to the standard cosmological phase. Then these modes are never  able to exit the horizon, and thus  do not undergo squeezing to produce coherent perturbations at late times.  The exact threshold discriminating between these behaviours depends of course on the model parameters.

While the value of $\gamma$  is fixed by the scaling properties of the scalar power spectrum, see Section \ref{sec:Constraints}, that of $\gamma_T$ is unconstrained. However the exact value of $\gamma_T$  does not change the qualitative behaviour of Eq. \eqref{eq:etaketaend}, for either blue ($\gamma_T>2$) or red  ($\gamma_T<2$) power spectrum. So in the following we focus on $\gamma_T\simeq 2$ for simplicity. Note that in principle this choice affects the lower bound on $\lambda_T$.

Also the dependence of $\eta_k/\eta_{\text{end}}$ on the equation of state via the parameter $m$ is not very relevant. Indeed, only factors within brackets in Eq. \eqref{eq:etaketaend} discriminate between the two cases of $\eta_k/\eta_{\text{end}}$ being smaller or larger than unity.  In this context $m$ only provides a contribution of order unity (excluding the vicinity of $m=1$, i.e. $w=1/3$), which is negligible compared to the order-of-magnitude estimates we use for the other quantities in the brackets.

So the remaining relevant model parameter is $\lambda$ ($\lambda_T$ for tensor modes). From observations of scalar modes $\lambda\simeq 10^{-24}\,\text{eV}^{-1}$  (see Section \ref{sec:Constraints}). Concerning $\lambda_T$, as we mentioned in Section \ref{sec:Constraints} we can only draw a lower bound on its value, while  possible upper bounds derived from the direct observation of gravity waves from binary black holes are very weak.
With this in mind we can inspect the relation \eqref{eq:etaketaend}. For scalar modes we find that all modes with $k< 10^{-5}$ exit the horizon before the MDR phase ends. So all scalar modes in the observable range can produce the observed standing waves (see also \cite{Gubitosi:2017zoj, Gubitosi:2017jwi}).

For tensor modes the range of  modes that can exit the horizon depends on the value of $\lambda_T$. Calling $k_T$ the maximum wavenumber that  allows for horizon exit we find:
\be
k_T= \sqrt{\frac{H_0\sqrt{\Omega_{r,0}}}{\lambda_T} \sqrt{\frac{|m-1|}{m}}}\simeq 10^{-17}\lambda_T^{-1/2}\label{eq:kMax}
\ee

Since tensor modes have not been observed yet we can not use this relation  to constrain the scale of deformation of the dispersion relation of gravity waves. We can however take a complementary perspective and ask what values of $\lambda_T$ could produce nontrivial signatures.
For $k_T$ to fall within the CMB observable range $\lambda_T$ would need to have a macroscopically large value, thus ruling this possibility out. 
On the other hand, the space interferometer LISA will be mostly sensitive to gravity waves in the frequency range  $(10^{-4} - 10^{0}) \, \text{Hz}$, corresponding to wavenumbers $k\sim(10^{-18}-10^{-13})\,\text{eV}$. Tensor perturbations in such range would not produce standing waves if $\lambda_T\gtrsim 10^{-8}\,\text{eV}^{-1}$, a value that, while being compatible with current limits from black hole binaries, is still unrealistically large.
Matters could become more interesting if ground-based interferometers were to observe a primordial signal, since they  are  sensitive to $k\sim(10^{-13}-10^{-9})\,\text{eV}$ \cite{Cabass:2015jwe}, and tensor modes in this range would not produce coherent perturbations for $\lambda_T\gtrsim 10^{-16}\text{eV}^{-1}$, i.e. if the deformation scale is in the PeV range. 
Whether or not such signal is actually observable depends on the amplitude of tensor modes in the relevant frequency  range. At the moment no explicit computation of the predicted amplitude of tensor modes in MDR models is available in the literature, except for the rough estimate discussed at the end of Section \ref{sec:Constraints} under quite stringent assumptions. Since the issue of detectability of primordial tensor modes with space- or ground-based interferometers might prove crucial in distinguishing MDR scenarios from the standard inflationary one, we plan to address it in  future work.

\section{Conclusions}
In this paper we have re-examined the MDR solution to the primordial fluctuations problem, bringing together
(and correcting) a number of results, confirming in detail some previous findings, and deriving one new striking result. 

We first revisited the
condition for a solution to the horizon problem, which, we recall, is a necessary condition for any primordial structure formation 
scenario. We corrected a mistake that crept into most of the earlier literature, where it was erroneously implied
that inflation and MDR were incompatible. Here we derived the correct condition between the UV form of the dispersion relations
and the equation of state $w$, explicitly showing that accelerated expansion and MDR scenarios may co-exist~\cite{Bianco:2016yib}.
Whether such scenarios go against Occam's razor, or instead could prove crucial in relating this work
with string theory and quantum gravity, remains to be seen.

Focussing on the ``MDR \emph{without} inflation'' scenario, we then examined in detail the general issue which for  inflationary scenarios goes under the tag  of  ``squeezing''. 
Cosmological fluctuations are typically produced as travelling waves when they first see the light of
day, deep inside the horizon. They then leave the horizon, under whatever mechanism resolved the horizon problem
of Big Bang cosmology, to re-enter the horizon much later as standing waves. The mechanism behind this conversion was examined 
in~\cite{Gubitosi:2017zoj,Gubitosi:2017jwi} for alternative scenarios. ``Quantum'' squeezing in inflation was  found
to be nothing but the prevalence of growing modes over decaying modes in more general scenarios. A case 
was left opened: MDR models with a blue spectrum. Indeed, in  \cite{Gubitosi:2017zoj, Gubitosi:2017jwi} we derived the conditions that  in the MDR scenario allow for  modes to produce standing waves once they exit the horizon. This entails that the modes spend a long enough time outside the horizon in order for the momentum of the perturbations to become subdominant. We showed that if the spectrum of perturbations is red (matching observed properties of scalar perturbations) then this condition is realised regardless of the equation of state while modes are outside the horizon \cite{Gubitosi:2017zoj}. When perturbations have a blue spectrum, as might be the case for the  as yet unobserved tensor modes, they might or might not produce standing waves with the correct temporal phase, depending on the equation of state while modes are outside the horizon.  In particular, perturbations produce standing waves with a cosine phase if $w>1$ and travelling waves if $w=1$, while $w<1$ reproduces the usual standing waves with a sine phase \cite{Gubitosi:2017jwi}. 

In this paper we took this analysis one step further, focussing on whether perturbations are actually able to exit the horizon at all before the MDR phase ends. We found  that the qualitative picture does not depend on $\gamma$ nor on the equation of state during the MDR phase.  What matters is the  scale of deformation, $\lambda$ and $\lambda_T$ for scalar and tensor modes respectively. This scale directly affects the maximum value of the wavenumber of a mode that can exit the horizon before ending the MDR phase, as stated in Eq. \eqref{eq:kMax}. Given the observational constraint on the scale of deformation of scalar modes $\lambda$, we found that these modes are always allowed to exit the horizon, unless the wavenumber takes a value that is way too large to be relevant for cosmological observations. This   result confirms once again that MDR models are compatible with observationally viable scalar perturbations, thus strengthening the case for their relevance in cosmology.

The situation is somewhat different for tensor modes, because their scale of deformation is still unconstrained, since they have yet to be observed. In addition, the range of wavelengths that could in principle be observed by current and future observations is much larger that that of scalar modes, because of the possibility of direct observation by gravity-waves interferometers. We found that for values of the deformation parameter comparable to the scalar one, all relevant modes would still be able to exit the horizon (then whether or not they produce standing waves depends on the factors discussed in \cite{Gubitosi:2017jwi}). Matters would change were the tensor modes sensitive to a deformation scale at lower energies, around the PeV range. In this case modes falling in the range testable by ground based interferometers would have always lived inside the horizon, preventing the formation of coherent standing waves. Such low-energy deformation parameter might seem unnatural from the point of view of quantum gravity but such scenarios have been entertained, starting from~\cite{ArkaniHamed:1998rs}. In addition we emphasize that such scenarios are still very much admissible given current constraints. In fact, for the specific purposes of this paper (and ignoring its contribution), the only available information so far comes from measurements of the speed of gravity waves from binary black holes mergers, which can however only put a lower limit well below the eV range \cite{Sotiriou:2017obf}.

Whatever the case our paper makes predictions that sooner or later will be relevant to gravitational wave experiments, should there be a primordial background of gravitational waves. 

In closing, we note that 
when it comes to the important issue of {\it predictivity}~\cite{Gubitosi:2015pba}, it must be admitted that MDR models, in their current state of development and relation to quantum gravity, do not fare better than inflation (and certainly fare much worse than bimetric models~\cite{Afshordi:2016guo}). But such considerations ignore predictions beyond the power spectrum parameters for scalar and tensor modes (although even at that level it is possible that the situation might change). The prospect that what for all other scenarios appears as standing waves becomes  travelling waves in MDR models could be a striking tell-tale signature. 
The only caveat may be, of course, that the trans-Planckian problem of inflation would be resolved in a manner that leads to similar predictions~\cite{Contaldi:2018akz}. Once again, inflation falls victim of its pliability. 
The issue of detectability with reference to the standing/stationary nature of the waves has been examined in~\cite{Renzini:2018vkx,Contaldi:2018akz}.

\section{Acknowledgements}
We thank Carlo Contaldi and Arianna Renzini for discussions related to this paper.   JM was supported by  an STFC consolidated  grant. GG acknowledges support from the Junta de Castilla y Le\'on (Spain) under grant BU229P18.

\end{document}